\documentclass[
reprint,
superscriptaddress,
frontmatterverbose, 
showpacs,
preprintnumbers,
nofootinbib,
amsmath,
amssymb,
aps,
floatfix,
twocolumn
]{revtex4-2}

\usepackage{graphicx}
\usepackage{dcolumn}
\usepackage{bm}
\usepackage{hyperref}
\usepackage{xcolor}
\usepackage{subcaption}
\usepackage{booktabs,array}

\begin{document}


\title{
Asymmetric Inelastic Dark Matter and the  LUX-ZEPLIN event
}

\author{Natsumi~Nagata}
 \email{natsumi@hep-th.phys.s.u-tokyo.ac.jp}
  \affiliation{Department of Physics, University of Tokyo, Bunkyo-ku, Tokyo
 113--0033, Japan
 }
\author{Tsutomu T. Yanagida}
 \email{tsutomu.tyanagida@ipmu.jp}
 \affiliation{Kavli Institute for the Physics and Mathematics of the Universe (WPI), UTIAS, The University of Tokyo, Chiba 277-8583, Japan
}
 \affiliation{Tsung-Dao Lee Institute \& School of Physics and Astronomy, Shanghai Jiao Tong University, Pudong New Area, Shanghai 201210, China
 }

\date{\today}

\begin{abstract}

  The recent LUX-ZEPLIN (LZ) search at high nuclear-recoil energies reported an intriguing candidate event near \(250~\mathrm{keV}\), motivating interpretations in terms of inelastic dark matter. A simple realization with Higgsino dark matter, however, is strongly constrained by capture and subsequent annihilation in the Sun, which would produce a high-energy neutrino flux excluded by IceCube. We propose an asymmetric inelastic dark-matter scenario that naturally avoids this tension. The dark matter is a predominantly Standard-Model-singlet Dirac fermion that mixes weakly with a nearly degenerate electroweak doublet. Its primordial asymmetry is generated by CP-violating out-of-equilibrium decays of heavy Majorana fermions produced nonthermally from inflaton decays, while the symmetric component is efficiently depleted through dark-sector interactions. The absence of an appreciable antiparticle abundance then suppresses dark-matter annihilation in the Sun and eliminates the associated IceCube constraint. The singlet--doublet structure simultaneously yields an off-diagonal \(Z\)-boson coupling proportional to the mixing angle, whereas the diagonal coupling is suppressed by its square. Consequently, an observable inelastic scattering rate can coexist with stringent limits on elastic scattering. We find that the LZ event can be accommodated for dark-matter masses of a few hundred GeV to \(1~\mathrm{TeV}\), with mass splittings of a few hundred keV and mixing angles below \(\mathcal O(10^{-2})\). Since the relic abundance is set by the primordial asymmetry rather than thermal freeze-out, lighter dark matter is also viable and can alleviate the potential tension with the absence of events in the LZ high-energy sideband. This framework predicts correlated elastic and inelastic direct-detection signals, together with complementary signatures at collider and precision experiments.

\end{abstract}

\maketitle


\section{Introduction}
\label{sec:introduction}

Recently, the LUX-ZEPLIN (LZ) Collaboration reported an intriguing high-energy nuclear-recoil candidate~\cite{LZ:2026axp}.  Using an exposure of \(2.84~{\rm tonne\mbox{-}yr}\), LZ extended the nuclear-recoil search window up to approximately \(270~{\rm keV}\), thereby gaining sensitivity to DM scenarios in which the recoil spectrum is shifted toward higher energies, such as inelastic dark matter.  One event was observed with characteristics consistent with a nuclear recoil of \(E_R = 248\pm23~({\rm stat})\pm23~({\rm sys})~{\rm keV}\), in a region where the expected background is small.  The maximum local significance among the interaction hypotheses considered by LZ reaches \(3.4\sigma\), while the global significance after accounting for the look-elsewhere effect is \(2.6\sigma\).  Although this event by itself does not constitute evidence for DM, its unusually large recoil energy and the absence of a clear background explanation\footnote{{For a discussion of possible Standard Model background contributions to the LZ event, including uncertainties associated with enhanced recombination in xenon, see Ref.~\cite{LZ:2025hud}.}} make it interesting to explore possible particle-physics interpretations.

A particularly simple interpretation of the LZ event was promptly proposed in terms of Higgsino dark matter~\cite{Fan:2026kxx,Freese:2026sga,Wu:2026nhi}.  A nearly pure neutral Higgsino consists of two nearly degenerate Majorana states, \(\widetilde{\chi}_1^0\) and \(\widetilde{\chi}_2^0\),\footnote{For phenomenology of such a Higgsino, see Ref.~\cite{Nagata:2014wma}.} whose coupling to the \(Z\) boson is predominantly off diagonal.  The lighter state can therefore scatter endothermically off a xenon nucleus \(N\), \( \widetilde{\chi}_1^0 + N \to \widetilde{\chi}_2^0 + N\), with the mass splitting \(\delta_{\widetilde H}\equiv m_{\widetilde{\chi}_2^0}-m_{\widetilde{\chi}_1^0}\) shifting the recoil spectrum toward higher energies.\footnote{The same inelastic-scattering mechanism can also arise for more general hypercharged fermionic dark matter split into two nearly degenerate Majorana states~\cite{Nagata:2014aoa}.} For a thermal Higgsino whose relic abundance accounts for all of the DM, \(m_{\widetilde H}\simeq1.1~{\rm TeV}\)~\cite{Cirelli:2007xd}, the LZ event can be accommodated for a mass splitting of a few hundred keV.  In particular, \(\delta_{\widetilde H}\simeq350~{\rm keV}\) provides a representative value for the Standard Halo Model, although the preferred splitting depends sensitively on the high-velocity tail of the Galactic DM distribution~\cite{Fan:2026kxx,Freese:2026sga,Rodd:2026tyn}.  Subsequent studies have explored the supersymmetric realization and complementary phenomenology of this interpretation~\cite{Du:2026guj,Bisal:2026khf,Cheung:2026byg,Langhoff:2026ujr, Frolovsky:2026tvq}.

This interpretation, however, is strongly constrained by the capture of Higgsino DM in the Sun~\cite{Pospelov:2026ewn}.  Owing to the deep gravitational potential of the Sun, incoming halo DM is accelerated to velocities substantially larger than those relevant for terrestrial direct-detection experiments.  This allows endothermic scattering on heavy solar nuclei, and hence efficient capture, even for mass splittings for which scattering on terrestrial targets is strongly suppressed~\cite{Pospelov:2026ewn}. The captured Higgsinos can subsequently annihilate predominantly into electroweak gauge bosons, in particular \(W^+W^-\) and \(ZZ\), producing high-energy neutrinos that may be searched for with neutrino telescopes.  IceCube has recently placed stringent limits on high-energy neutrinos from DM annihilation in the Sun using $\simeq 10$ years of data~\cite{IceCube:2025fcu}.  Applying these limits to the thermal Higgsino scenario, Ref.~\cite{Pospelov:2026ewn} obtains \(\delta_{\widetilde H}>566~{\rm keV}\), which is incompatible with the mass splitting of a few hundred keV preferred by the LZ event for a standard thermal Higgsino.  Subsequent independent analyses have reached similar conclusions, finding lower limits on the Higgsino mass splitting of approximately \(0.5\)--\(0.6~{\rm MeV}\)~\cite{DiMauro:2026dqp,Bose:2026ndd, Nguyen:2026lui}.\footnote{This IceCube limit can be evaded by considering a much heavier Higgsino~\cite{Langhoff:2026ujr}, with a mass of order \(10^5\)--\(10^6\,\mathrm{GeV}\). For a viable scenario realizing such heavy Higgsino dark matter, see Ref.~\cite{Feldstein:2013uha}.}

Motivated by this tension, we propose a simple framework in which the very feature that suppresses the solar-neutrino signal—an asymmetric dark-matter population—coexists naturally with the inelastic scattering required to account for the LZ high-recoil event. In an asymmetric dark-matter scenario~\cite{Zurek:2013wia}, the antiparticle abundance is strongly depleted, so that dark matter captured in the Sun has essentially no annihilation partner~\cite{Frandsen:2010yj} and therefore does not produce the high-energy neutrino flux constrained by IceCube. The dark-sector asymmetry is generated through a mechanism closely analogous to nonthermal leptogenesis~\cite{Lazarides:1990huy,Kumekawa:1994gx,Asaka:1999yd,Asaka:1999jb}: inflaton decays first produce heavy singlet Majorana fermions, whose subsequent CP-violating out-of-equilibrium decays generate asymmetries in a scalar and a fermion in the dark sector. Each decay produces one scalar and one fermion, locking their number asymmetries together; crucially, however, these two states are distinct species rather than particle--antiparticle partners, so their equal production does not constitute a symmetric population that can annihilate away. At low energies, the predominantly SM-singlet fermionic dark matter mixes weakly with a nearly degenerate electroweak-doublet fermion. This structure allows the elastic and inelastic direct-detection rates to be parametrically separated: a small singlet--doublet mixing suppresses elastic scattering, while retaining a sufficiently large \(Z\)-mediated transition between the nearly degenerate states to generate the high-energy nuclear recoil observed by LZ. The resulting framework thus provides an economical way to reconcile an inelastic interpretation of the LZ event with the stringent solar-neutrino constraints from IceCube.

The remainder of this paper is organized as follows. In Sec.~\ref{sec:model}, we introduce the model and discuss the mass spectrum and interactions of the dark-sector particles, with particular emphasis on the singlet--doublet mixing that controls their couplings to the \(Z\) boson. In Sec.~\ref{sec:asymmetry}, we describe the nonthermal generation of the dark-sector asymmetry through the CP-violating decays of heavy Majorana fermions and its subsequent transfer to the dark fermions. In Sec.~\ref{sec:depletion}, we study the cosmological evolution of the dark sector, including the depletion of the symmetric component, the resulting dark-matter abundance, and the decays of the heavier dark-sector states. In Sec.~\ref{sec:dark-photon-elastic}, we examine elastic dark-matter scattering mediated by dark-photon exchange and the associated direct-detection constraints. In Sec.~\ref{sec:lz}, we turn to the \(Z\)-mediated elastic and inelastic scattering processes and show that the LZ high-recoil event can be accommodated while satisfying existing elastic-scattering bounds. We also discuss characteristic predictions of the scenario and its distinction from other singlet--doublet interpretations of the LZ event. Finally, Sec.~\ref{sec:discussion} is devoted to a summary and discussion of our results.

\section{Model}
\label{sec:model}

We extend the Standard Model (SM) gauge group by an Abelian factor \(\mathrm{U}(1)_D\), with gauge coupling \(g_D\) and dark gauge boson \(\gamma_D\). Throughout this paper, all fermionic fields are written as left-handed two-component Weyl spinors. The new field content consists of complex scalars \(S\) and \(\Phi\), a vector-like pair of SM-singlet fermions \(D\) and \(D^c\), heavy gauge-singlet Majorana fermions \(N_i\), and a vector-like pair of electroweak-doublet fermions,
\begin{align}
\Psi &=
\begin{pmatrix}
\psi^+ \\
\psi^0
\end{pmatrix},
\qquad
\Psi^c =
\begin{pmatrix}
\psi^{c0} \\
\psi^-
\end{pmatrix}.
\end{align}
Their representations under \(G_{\rm SM}\times\mathrm{U}(1)_D\) are
\begin{align}
\Psi &: (\bm{1},\bm{2},+1/2)_{+1}\,, \nonumber\\
\Psi^c &: (\bm{1},\bm{2},-1/2)_{-1}\,,
\end{align}
where the subscripts denote the \(\mathrm{U}(1)_D\) charges. The scalars \(S\) and \(\Phi\) are SM singlets with dark charges \(-1\) and \(+3\), respectively. The fermions \(D\) and \(D^c\) have dark charges \(+1\) and \(-1\) and form a Dirac fermion, whereas the \(N_i\) are neutral under all gauge factors. The scalar \(\Phi\) spontaneously breaks \(\mathrm{U}(1)_D\).

The fields \(N_i\) are assumed to be distinct from the right-handed neutrinos \(N_R^c\) associated with the neutrino sector. This separation can be realized through a high-scale breaking of \(\mathrm{U}(1)_{B-L}\) that leaves the matter parity \(P_{B-L}=(-1)^{3(B-L)}\) unbroken~\cite{Krauss:1988zc, Ibanez:1991hv, Ibanez:1991pr, Martin:1992mq}. We assign vanishing \(B-L\) charge to \(N_i\), \(D\), \(D^c\), \(S\), \(\Phi\), \(\Psi\), and \(\Psi^c\), so that all these fields are even under \(P_{B-L}\). The SM matter fields and \(N_R^c\) are odd, while the Higgs doublet \(H\) is even. Matter parity therefore forbids the renormalizable operators \(N_iN_R^c\), \(N_iLH\), \(SDN_R^c\), and \(S^\dagger D^cN_R^c\), as well as the otherwise gauge-invariant coupling \(S\,\Psi\cdot L\), where \(A\cdot B\equiv\epsilon_{ab}A^aB^b\) with \(\epsilon_{12} = +1\).

The relevant quantum numbers are summarized in Table~\ref{tab:charges}.
\begin{table}[ht]
\centering
\renewcommand{\arraystretch}{1.2}
\begin{tabular}{c c c c c}
\toprule
Field & Spin & $G_{\rm SM}$ & $\mathrm{U}(1)_D$ & $P_{B-L}$ \\
\midrule
$N_i$    & $1/2$ & $(\bm{1},\bm{1},0)$    & $0$  & $+$ \\
$D$      & $1/2$ & $(\bm{1},\bm{1},0)$    & $+1$ & $+$ \\
$D^c$    & $1/2$ & $(\bm{1},\bm{1},0)$    & $-1$ & $+$ \\
$\Psi$   & $1/2$ & $(\bm{1},\bm{2},+1/2)$ & $+1$ & $+$ \\
$\Psi^c$ & $1/2$ & $(\bm{1},\bm{2},-1/2)$ & $-1$ & $+$ \\
$S$      & $0$   & $(\bm{1},\bm{1},0)$    & $-1$ & $+$ \\
$\Phi$   & $0$   & $(\bm{1},\bm{1},0)$    & $+3$ & $+$ \\
\midrule
$N_R^c$  & $1/2$ & $(\bm{1},\bm{1},0)$    & $0$  & $-$ \\
$L$      & $1/2$ & $(\bm{1},\bm{2},-1/2)$ & $0$  & $-$ \\
$H$      & $0$   & $(\bm{1},\bm{2},+1/2)$ & $0$  & $+$ \\
\bottomrule
\end{tabular}
\caption{Field content and charge assignments.}
\label{tab:charges}
\end{table}

In addition to the canonical kinetic terms, the Lagrangian contains
\begin{align}
\mathcal L \supset{}&-\biggl[
\frac12 M_{ij}N_iN_j+m_DDD^c
+m_\Psi\Psi^c\cdot\Psi \nonumber\\
&\quad +g_iN_iDS+\widetilde g_iN_iD^cS^\dagger \nonumber \\ 
&+yD\,\Psi^c\cdot H
+\widetilde yD^cH^\dagger\Psi
+\mathrm{h.c.}\biggr] \nonumber\\
&-V(H,S,\Phi)
-\frac{\epsilon}{2}B_{\mu\nu}F_D^{\mu\nu}\,,
\label{eq:massYuk}
\end{align}
where \(M_{ij}=M_{ji}\), and \(B_{\mu\nu}\) and \(F_D^{\mu\nu}\) are the field-strength tensors of the hypercharge gauge field  and the dark gauge field, respectively. The most general renormalizable scalar potential consistent with the stated gauge symmetries and matter parity is
\begin{align}
V(H,S,\Phi)&=m_S^2|S|^2+\lambda_S|S|^4
-\mu_\Phi^2|\Phi|^2+\lambda_\Phi|\Phi|^4 \nonumber\\
&+\lambda_{HS}|H|^2|S|^2
+\lambda_{H\Phi}|H|^2|\Phi|^2
+\lambda_{S\Phi}|S|^2|\Phi|^2 \nonumber\\
&+\bigl(\kappa\Phi S^3+\mathrm{h.c.}\bigr) + V_{\rm SM}(H)\,.
\label{eq:scalar-potential}
\end{align}
We choose the parameters such that the vacuum satisfies
\begin{align}
\langle H\rangle=\frac{1}{\sqrt2}\begin{pmatrix}0\\v\end{pmatrix},
\qquad
\langle\Phi\rangle=\frac{v_\Phi}{\sqrt2},
\qquad
\langle S\rangle=0.
\end{align}
The charge-three vacuum expectation value (VEV) $v_\Phi$ breaks \(\mathrm{U}(1)_D\) to a residual \(\mathbb Z_3\) gauge symmetry,\footnote{{This residual symmetry is distinct from \(P_{B-L}\), which also remains unbroken. In particular, it forbids Majorana masses for \(D\) and \(D^c\).}} which assures the stability of dark matter. Before accounting for neutral-gauge-boson mixing, the dark photon acquires the mass
\begin{align}
m_{\gamma_D}=3g_Dv_\Phi.
\label{eq:dark-photon-mass}
\end{align}

The kinetic mixing parameter \(\epsilon\),\footnote{Mixing between neutral vector fields was considered early on in the context of vector-meson dominance~\cite{Sakurai:1960ju} and was formulated in a gauge-invariant Lagrangian framework in Ref.~\cite{Kroll:1967it}.  Kinetic mixing between two Abelian gauge fields and its radiative generation in theories containing particles charged under both gauge groups were discussed in Ref.~\cite{Holdom:1985ag}; see also Ref.~\cite{Fabbrichesi:2020wbt} for further discussions and reviews.  For an analogous treatment of scalar kinetic mixing, see Ref.~\cite{Fukuda:1974kn}.} the Higgs portal couplings \(\lambda_{HS}\) and \(\lambda_{H\Phi}\), together with the electroweak interactions of the doublets and their Yukawa couplings, can establish thermal contact between the visible and dark sectors. 

After the electroweak symmetry breaking, with a Higgs VEV \(v \simeq 246~\mathrm{GeV}\), the neutral fermion mass terms can be written as
\begin{align}
\mathcal{L}_{\rm mass}^{(0)}
=
-
\begin{pmatrix}
D & \psi^0
\end{pmatrix}
\mathcal{M}_0
\begin{pmatrix}
D^c\\
\psi^{c0}
\end{pmatrix}
+\mathrm{h.c.},
\end{align}
where
\begin{align}
\mathcal{M}_0
=
\begin{pmatrix}
m_D & m_y\\
m_{\widetilde y} & m_\Psi
\end{pmatrix},
\quad
m_y\equiv\frac{yv}{\sqrt{2}},
\quad
m_{\widetilde y}\equiv\frac{\widetilde yv}{\sqrt{2}}.
\label{eq:neutral-mass-matrix}
\end{align}
We diagonalize the mass matrix by two unitary matrices \(U\) and \(V\),
\begin{align}
\begin{pmatrix}
D\\
\psi^0
\end{pmatrix}
=
U
\begin{pmatrix}
\chi_1\\
\chi_2
\end{pmatrix}
\,,
\quad 
\begin{pmatrix}
D^c\\
\psi^{c0}
\end{pmatrix}
=
V
\begin{pmatrix}
\chi_1^c\\
\chi_2^c
\end{pmatrix},
\end{align}
such that
\begin{align}
U^T\mathcal{M}_0V
=
\begin{pmatrix}
m_{\chi_1}&0\\
0&m_{\chi_2}
\end{pmatrix}
\,,
\quad
m_{\chi_1}<m_{\chi_2}\,,
\label{eq:neutral-diagonalization}
\end{align}
where the phases of the fields are chosen so that the physical masses are positive.  Each pair \((\chi_i,\chi_i^c)\) constitutes a four-component Dirac fermion, which we denote by
\begin{align}
\mathcal{X}_i\equiv
\begin{pmatrix}
\chi_i\\
\chi_i^{c\dagger}
\end{pmatrix}.
\end{align}

Since \(D\) and \(D^c\) are electroweak singlets, the coupling to the \(Z\) boson originates entirely from the neutral components of the doublets. In terms of the mass eigenstates, it is given by 
\begin{align}
\mathcal{L}_Z
=
\frac{g_2}{2c_W}Z_\mu
\biggl[
&
V_{2i}^*V_{2j}\,
\chi_i^{c\dagger}\bar{\sigma}^\mu\chi_j^c
-
U_{2i}^*U_{2j}\,
\chi_i^\dagger\bar{\sigma}^\mu\chi_j
\biggr],
\label{eq:Z-mass-basis}
\end{align}
where \(g_2\) is the \(\mathrm{SU}(2)_L\) gauge coupling constant, \(c_W\equiv \cos \theta_W\) with \(\theta_W\) the weak mixing angle, and summation over \(i,j=1,2\) is understood.  This expression contains both diagonal and off-diagonal interactions.  In particular, if the lighter state \(\mathcal{X}_1\) is predominantly singlet-like, its diagonal coupling to the \(Z\) boson is quadratic in the small doublet admixture, whereas the transition coupling between \(\mathcal{X}_1\) and \(\mathcal{X}_2\) is only linear in the mixing.  This hierarchy allows the \(Z\)-mediated inelastic transition \(\mathcal{X}_1+N\to\mathcal{X}_2+N\) to be sizable while the elastic \(Z\)-exchange amplitude of the lighter state is comparatively suppressed.

For illustration, consider the particularly simple case \(y=\widetilde y\), with all parameters real.  The mass matrix is then symmetric and can be diagonalized with a single mixing angle, \(U=V\), satisfying
\begin{align}
\tan2\theta
=
\frac{\sqrt{2}\,y v}{m_\Psi-m_D}\,.
\end{align}
The two masses are
\begin{align}
m_{\chi_{1,2}}
=
\frac{1}{2}
\left[
m_D+m_\Psi
\mp
\sqrt{(m_\Psi-m_D)^2+2y^2v^2}
\right]\,.
\label{eq:neutral-masses-symmetric}
\end{align}
In this case, the \(Z\)-boson interaction can be written in four-component notation as
\begin{align}
\mathcal{L}_Z
=
&-\frac{g_2}{2c_W}Z_\mu
\biggl[
\sin^2\theta\,
\overline{\mathcal{X}}_1\gamma^\mu\mathcal{X}_1
+
\cos^2\theta\,
\overline{\mathcal{X}}_2\gamma^\mu\mathcal{X}_2
\nonumber\\
&-
\sin\theta\cos\theta
\left(
\overline{\mathcal{X}}_1\gamma^\mu\mathcal{X}_2
+
\overline{\mathcal{X}}_2\gamma^\mu\mathcal{X}_1
\right)
\biggr].
\label{eq:Z-coupling-symmetric}
\end{align}
Thus, for \(\sin\theta\ll1\), the diagonal coupling of the singlet-like state scales as \(\sin^2\theta\), whereas the off-diagonal coupling scales as \(\sin\theta\).  This feature will play an important role in realizing an observable inelastic nuclear-recoil signal while suppressing conventional elastic direct-detection constraints.

In the following, we identify the lighter neutral Dirac fermion
\(\mathcal{X}_1\) as the dark matter (DM) candidate and consider
\(m_{\chi_1}\) in the range from a few hundred GeV to about
\(1~\mathrm{TeV}\). The heavier state \(\mathcal{X}_2\) is assumed
to be nearly degenerate with \(\mathcal{X}_1\), with a mass splitting
\begin{align}
\delta_\chi \equiv m_{\chi_2}-m_{\chi_1}
\simeq 300~\mathrm{keV}.
\label{eq:chi-mass-splitting}
\end{align}
In the simplified parameter choice discussed above, Eq.~\eqref{eq:neutral-masses-symmetric} gives \(\delta_\chi=\sqrt{(m_\Psi-m_D)^2+2y^2v^2}\). Thus, this small splitting requires both a small difference between the diagonal mass parameters and a small Yukawa-induced mixing term. For a predominantly singlet-like DM state, we take \(m_D<m_\Psi\) and \(\sqrt{2}|y|v\ll m_\Psi-m_D\), so that \(\delta_\chi\simeq m_\Psi-m_D\) and \(|\sin\theta|\ll1\), consistently with the coupling hierarchy described above. We further assume the mass hierarchy
\begin{align}
m_{\gamma_D} \ll m_{\chi_1} \simeq m_{\chi_2} \ll m_S \ll M_i \,,
\label{eq:dark-mass-hierarchy}
\end{align}
where the dark-photon mass is given by Eq.~\eqref{eq:dark-photon-mass}.\footnote{{Without loss of generality, we work in a basis in which the Majorana mass matrix is diagonal and real, \(M_{ij}=M_i\delta_{ij}\), with \(M_i>0\).}} The resulting spectrum thus consists of a light dark photon, two nearly degenerate fermions \(\chi_{1,2}\), a heavier scalar \(S\), and much heavier Majorana fermions \(N_i\).

\section{Asymmetry in the dark sector}
\label{sec:asymmetry}

Next, we discuss the generation and subsequent evolution of the particle--antiparticle asymmetry in the dark sector. The underlying mechanism is closely analogous to leptogenesis~\cite{Fukugita:1986hr}.\footnote{{For reviews of leptogenesis, see, e.g., Refs.~\cite{Buchmuller:2005eh, Davidson:2008bu, Fong:2012buy}.}} Related scenarios in which the decays of heavy right-handed neutrinos generate a dark-sector asymmetry have been studied in Refs.~\cite{Falkowski:2011xh,Josse-Michaux:2011sjn,Narendra:2018vfw}.

In our scenario, we assume that both the heavy Majorana fermions \(N_i\) and the right-handed neutrinos in the visible sector are produced non-thermally through inflaton decays. The \(N_i\) subsequently decay into dark-sector matter fields through the Yukawa couplings \(g_i\) and \(\widetilde g_i\) in Eq.~\eqref{eq:massYuk}. Since the \(N_i\) are Majorana fermions, they can decay into both particle and antiparticle final states. Interference between the tree-level and loop amplitudes gives rise to CP-violating differences between the corresponding decay rates, thereby generating a particle--antiparticle asymmetry in the dark sector. This mechanism may be viewed as a dark-sector analogue of standard non-thermal leptogenesis~\cite{Lazarides:1990huy,Kumekawa:1994gx,Asaka:1999yd,Asaka:1999jb}, in which heavy Majorana right-handed neutrinos are produced through inflaton decays and their subsequent CP-violating decays generate a cosmological asymmetry. In the present setup, the scalar \(S\) produced in \(N_i\) decays subsequently decays into two dark fermions through the dimension-five operators, transferring the asymmetry initially stored in the scalar sector to the fermionic sector and ultimately leaving a nonzero dark-fermion asymmetry.

\subsection{Production of correlated fermion and scalar asymmetries}
\label{subsec:asym-production}

For later convenience, we define the CP asymmetry per $N_i$ decay as the net excess of Dirac particles, equivalently of $S$ particles, produced in the decay:
\begin{align}
\epsilon_i
\equiv
\frac{1}{\Gamma_i}
\biggl[
&
\Gamma(N_i\to DS)
+\Gamma(N_i\to D^{c\dagger}S)
\nonumber\\
&-\Gamma(N_i\to D^\dagger S^\dagger)
-\Gamma(N_i\to D^cS^\dagger)
\biggr]\,,
\label{eq:epsilon-def}
\end{align}
where \(\Gamma_i\) denotes the total tree-level decay width of \(N_i\),
\begin{align}
\Gamma_i
\equiv
\frac{M_i}{16\pi}
\left(
|g_i|^2+|\widetilde g_i|^2
\right).
\label{eq:Ni-width}
\end{align}
The asymmetry vanishes at tree level and is generated by the interference of the tree-level amplitudes with one-loop vertex and self-energy corrections. A nonzero asymmetry therefore requires at least two heavy Majorana fermions and irreducible CP-violating phases in their Yukawa couplings. In particular, the relevant CP-odd combinations include \(\operatorname{Im}\left[(g_i^*g_j)^2\right]\) and \(\operatorname{Im}\left[(\widetilde g_i^*\widetilde g_j)^2\right]\), whose nonzero values give rise to the dark-sector asymmetry.

We define the asymmetry yields by
\begin{align}
Y_{\Delta F}
&\equiv \sum_a \frac{n_{\Psi_a}-n_{\overline{\Psi}_a}}{s}\,,
\quad
Y_{\Delta S}
\equiv \frac{n_S-n_{S^\dagger}}{s}\,,
\label{eq:ydels}
\end{align}
where $s$ is the entropy density and $\Psi_a$ denotes the singlet and doublet Dirac fermions, or their mass eigenstates after the electroweak symmetry breaking. The decays of $N_i$ generate equal asymmetries in the fermion and scalar sectors, \(Y_{\Delta F}^{\mathrm{gen}} =Y_{\Delta S}^{\mathrm{gen}} \equiv Y_\Delta^{\mathrm{gen}}\). The particle-number asymmetries have the same sign, whereas their dark gauge charges cancel: \(Y^{\mathrm{gen}}_{Q_D}=Y^{\mathrm{gen}}_{\Delta F}-Y^{\mathrm{gen}}_{\Delta S}=0\) at production.

We denote the inflaton by $\phi$, its mass by $m_\phi$, and the reheating temperature by $T_R$. We do not specify the inflation sector in detail, but parameterize the production of the heavy Majorana fermions by the inflaton branching fractions, \(B_i^\phi \equiv {\rm Br}(\phi\to N_iN_i)\). The pair-production channel is kinematically allowed for \(m_\phi>2M_i\). As mentioned above, we also assume that the inflaton decays into right-handed neutrinos in the SM sector, thereby reheating the visible sector and generating the baryon asymmetry of the Universe through non-thermal leptogenesis.

The inflaton decays when the Hubble expansion rate \(H\) becomes approximately equal to the inflaton decay width \(\Gamma_\phi\): $H\simeq\Gamma_\phi$. We define the reheating temperature by
\begin{align}
\Gamma_\phi
=
H(T_R),
\qquad
H(T)
=
1.66\sqrt{g_*}\frac{T^2}{M_{\rm Pl}},
\label{eq:TR-definition}
\end{align}
where $M_{\rm Pl}=1.22\times10^{19}\ {\rm GeV}$ and \(g_*\) counts the total number of effective massless degrees of freedom. In the instantaneous-reheating approximation, the inflaton abundance at reheating is given by
\begin{align}
\frac{n_\phi}{s}
\simeq
\frac{3}{4}\frac{T_R}{m_\phi}\,,
\end{align}
where \(n_\phi\) denotes the inflaton number density. 

For simplicity, in what follows we focus on the regime \(\Gamma_i \gg \Gamma_\phi\), in which the $N_i$ decay promptly after being produced by the inflaton. Since each $\phi\to N_iN_i$ decay produces two $N_i$, their non-thermal yield is
\begin{align}
Y_{N_i}^{\rm NT}
\simeq
\frac{3}{2}
B_i^\phi
\frac{T_R}{m_\phi}\,,
\label{eq:YN-nonthermal}
\end{align}
where \(B_i^\phi = {\rm Br}(\phi\to N_iN_i)\) is the branching fraction of the inflaton decaying into \(N_i\). We assume \(T_R \ll M_i\), so that the inverse-decay processes \(DS \to N_i\) and \(D^c S^\dagger \to N_i\) are exponentially suppressed and the $N_i$ decays remain effectively out of equilibrium.\footnote{We note, however, that the maximum temperature \(T_{\mathrm{max}}\) attained during perturbative reheating can be significantly higher than the conventional reheating temperature~\cite{Chung:1998rq,Giudice:2000ex}. If \(T_{\mathrm{max}}\) is relevant for the thermalization of the dark sector, the condition above should be imposed using \(T_{\mathrm{max}}\) rather than \(T_R\). The same replacement should be made when evaluating the washout condition in Eq.~\eqref{eq:DeltaX-condition}.} We thus have 
\begin{align}
  Y_\Delta^{\mathrm{gen}} \equiv Y_{\Delta F}^{\mathrm{gen}} =Y_{\Delta S}^{\mathrm{gen}} \simeq
\frac{3}{2}
\frac{T_R}{m_\phi} \sum_i \epsilon_i B_i^\phi \,.
\end{align}

Below the mass scale of the heavy Majorana fermions, integrating out the $N_i$ generates the dimension-five operators
\begin{align}
{\cal L}_{\rm eff}
\supset
&
\frac{1}{2}C_g(DS)(DS)
+
\frac{1}{2}C_{\widetilde g}
(D^cS^\dagger)(D^cS^\dagger)
\nonumber\\
&+
C_0(DS)(D^cS^\dagger)
+\text{h.c.},
\label{eq:effective-washout}
\end{align}
where
\begin{align}
C_g
&=
\sum_i\frac{g_i^2}{M_i},
&
C_{\widetilde g}
&=
\sum_i\frac{\widetilde g_i^2}{M_i},
&
C_0
&=
\sum_i\frac{g_i\widetilde g_i}{M_i}.
\label{eq:C-coefficients}
\end{align}
The first two operators induce washout processes such as \(DS \leftrightarrow D^\dagger S^\dagger\) and \(DD \leftrightarrow S^\dagger S^\dagger\), together with the corresponding processes involving $D^c$. By contrast, the mixed operator proportional to $C_0$ does not erase the net asymmetry, although it can redistribute the asymmetry among the different dark-sector components.

The rate of the processes induced by these effective operators scales as
\begin{align}
\Gamma_{\Delta X}
\simeq
c_{\Delta X}
T^3
\left(
|C_g|^2+|C_{\widetilde g}|^2
\right),
\label{eq:DeltaX-rate}
\end{align}
where $c_{\Delta X}$ is a numerical coefficient of order $10^{-2}$--$10^{-1}$. Since $\Gamma_{\Delta X}/H\propto T$ during radiation domination, the washout constraint is strongest at the highest temperature attained after the asymmetry has been generated. Requiring the above processes to remain out of equilibrium therefore gives
\begin{align}
c_{\Delta X}T_R^3
\left(
|C_g|^2+|C_{\widetilde g}|^2
\right)
<
1.66\sqrt{g_*}\frac{T_R^2}{M_{\rm Pl}}.
\label{eq:DeltaX-condition}
\end{align}
Equivalently,
\begin{align}
&\sqrt{|C_g|^2+|C_{\widetilde g}|^2}
\lesssim
1.2\times10^{-12}\ {\rm GeV}^{-1}
\nonumber\\
&\times
\biggl(\frac{c_{\Delta X}}{10^{-2}}\biggr)^{-1/2}
\left(\frac{g_*}{106.75}\right)^{1/4}
\left(\frac{T_R}{10^8\ {\rm GeV}}\right)^{-1/2}.
\label{eq:DeltaX-bound}
\end{align}
This condition is readily satisfied over a broad range of parameter space. For example, couplings of order \(g_i,\widetilde g_i=\mathcal{O}(0.1)\) are compatible with the washout bound for heavy-fermion masses of order \(10^{10}\)--\(10^{11}\ {\rm GeV}\), depending on the number of contributing states and possible cancellations among the terms in Eq.~\eqref{eq:C-coefficients}.

\subsection{Scalar decay and asymmetry transfer}
\label{subsec:asym-transfer}

The asymmetry generated by the non-thermal \(N_i\) decays is initially shared between the fermions and the scalar \(S\). In the present scenario, however, \(S\) is much heavier than the dark fermions and does not survive as a stable relic. It can decay through the gauge-invariant dimension-five operators
\begin{align}
 \mathcal L_{\mathrm{dec}}^{(5)}
 =-\biggl[
 \frac{c_L}{2\Lambda}\Phi^\dagger S^\dagger DD
 +\frac{c_R}{2\Lambda}\Phi S D^cD^c
 +\mathrm{h.c.}\biggr],
 \label{eq:asym-dim5}
\end{align}
where \(c_L\) and \(c_R\) are dimensionless coefficients and
\(\Lambda\) parameterizes the heavy scale suppressing the effective interactions.

These interactions can arise radiatively from the renormalizable
couplings already present in the model. Combining
\(\kappa^*\Phi^\dagger S^{\dagger3}\) with the operator
\(C_g(DS)(DS)/2\) in Eq.~\eqref{eq:effective-washout}, and
contracting two scalar lines, generates the first operator in
Eq.~\eqref{eq:asym-dim5} at one loop. The conjugate scalar vertex
and the operator proportional to \(C_{\widetilde g}\) generate the
second. Parametrically, the induced coefficients scale as
\begin{align}
 \frac{c_L}{\Lambda}&\sim
 \frac{\kappa^*}{16\pi^2}
 \sum_i\frac{g_i^2}{M_i} \,, \quad 
 \frac{c_R}{\Lambda}\sim
 \frac{\kappa}{16\pi^2}
 \sum_i\frac{\widetilde g_i^2}{M_i}\,.
 \label{eq:transfer-loop-scaling}
\end{align}
Additional ultraviolet contributions can be included in the total coefficients \(c_{L,R}/\Lambda\).

The operators in Eq.~\eqref{eq:asym-dim5} induce the decay processes 
\begin{align}
 S&\longrightarrow DD \Phi^\dagger \,,
 \quad D^{c\dagger}D^{c\dagger} \Phi^\dagger \,,
 \nonumber\\
 S^\dagger&\longrightarrow D^\dagger D^\dagger \Phi \,,
 \quad D^cD^c \Phi \,.
 \label{eq:asym-S-decays}
\end{align}
Both channels of \(S\) produce two Dirac particles, whereas both channels of \(S^\dagger\) produce two antiparticles; the two terms do not transfer asymmetries of opposite sign. The widths of these three-body decay processes scale as
\begin{align}
 \Gamma_{S}\sim
 \frac{(|c_L|^2+|c_R|^2)m_{S}^3}
 {(4\pi)^3\Lambda^2},
 \label{eq:asym-three-body-scaling}
\end{align}
up to numerical and phase-space factors. The decay occurs around 
\begin{align}
  T_{S, \mathrm{decay}} &\simeq \biggl(\frac{(|c_L|^2+|c_R|^2) M_{\mathrm{Pl}}m_{S}^3}{1.66(4\pi)^3\Lambda^2 \sqrt{g_*}}\biggr)^{\frac{1}{2}} \nonumber \\ 
  &\simeq 2\times 10^4~\mathrm{GeV} \times \sqrt{|c_L|^2+|c_R|^2} \nonumber \\ 
  &\times  \biggl(\frac{g_*}{106.75}\biggr)^{-\frac{1}{4}} \biggl(\frac{m_S}{10^6~\mathrm{GeV}}\biggr)^{\frac{3}{2}}  \biggl( \frac{\Lambda}{10^{12}~\mathrm{GeV}}\biggr)^{-1} \,.
\end{align}
For suitable parameters, scalar decay can occur while the dark fermions are still relativistic.

We note, however, that the scalar asymmetry can be reduced before
decay by scalar-number-changing reactions induced by the coupling
\(\kappa\) in Eq.~\eqref{eq:scalar-potential}, such as
\(SS\leftrightarrow S^\dagger\Phi^\dagger\).
If these reactions remain efficient as \(S\) becomes nonrelativistic,
the scalar asymmetry is substantially suppressed. In this case,
the final fermion asymmetry is  given by
\begin{equation}
Y_{\Delta F}^{\mathrm{final}}
\simeq Y_{\Delta F}^{\mathrm{gen}}\,.
\label{eq:ydelf}
\end{equation}

If, on the other hand, these scalar-number-changing reactions are
ineffective because \(\kappa\) is very small, the final fermion
asymmetry is
\begin{align}
Y_{\Delta F}^{\mathrm{final}}
\simeq 3Y_{\Delta F}^{\mathrm{gen}}\,,
\end{align}
since each excess fermion produced in \(N_i\) decay is accompanied
by an excess \(S\) particle, whose subsequent decay produces two
additional fermions.

Given this uncertainty, we adopt Eq.~\eqref{eq:ydelf} in the following
analysis to obtain a conservative estimate of the DM abundance.

\section{Depletion of symmetric components}
\label{sec:depletion}

We next discuss the depletion of the symmetric dark-fermion population and the resulting DM abundance. In the present scenario, the heavy scalar \(S\) decays and does not contribute to the present relic density. Its decays nevertheless inject both symmetric and asymmetric fermion populations. We focus on parameters for which these decays are completed before dark-fermion freeze-out, and for which the decay products thermalize. The Higgs portals $\lambda_{HS}|H|^2|S|^2$ and $\lambda_{H\Phi}|H|^2|\Phi|^2$, kinetic mixing, and interactions of the electroweak doublets maintain a common dark and visible temperature during the relevant depletion epoch, while dark gauge interactions establish kinetic equilibrium within the dark sector. 

\subsection{Annihilation and coannihilation}
\label{subsec:depletion-annihilation}

All neutral Dirac mass eigenstates retain the same diagonal dark gauge coupling, since the singlet and doublet have equal dark charge. In particular, \(\mathcal X_i\overline{\mathcal X}_i\to \gamma_D\gamma_D\) is not suppressed by the singlet--doublet mixing. For \(m_{\gamma_D}\ll m_{\chi_i}\), the nonrelativistic Born cross section is
\begin{align}
 \left\langle\sigma v_{\rm rel}\right\rangle_
 {\mathcal X_i\overline{\mathcal X}_i\to\gamma_D\gamma_D}
 \simeq\frac{\pi\alpha_D^2}{m_{\chi_i}^2}\,,
 \label{eq:depletion-dark-annihilation}
\end{align}
with \(\alpha_D \equiv g_D^2/(4\pi)\). Its numerical value
at the common neutral mass \(m\equiv m_{\chi_1} \simeq m_{\chi_2} \) is
\begin{align}
   \left\langle\sigma v_{\rm rel}\right\rangle_
 {\mathcal X_i\overline{\mathcal X}_i\to\gamma_D\gamma_D}
 &\simeq 1\times10^{-25}~\mathrm{cm^3\,s^{-1}} \nonumber \\ 
 &\times 
 \left(\frac{\alpha_D}{0.03}\right)^2
 \left(\frac{500~\mathrm{GeV}}{m}\right)^2.
 \label{eq:depletion-dark-rate}
\end{align}

The doublet components also annihilate into electroweak gauge bosons, including \(W^+W^-\) and \(ZZ\). Charged-pair annihilation and neutral--charged coannihilation \cite{Griest:1990kh} add channels such as \(\gamma\gamma\), \(\gamma Z\), \(W^\pm Z\), and \(W^\pm\gamma\), with the appropriate initial electric charges. For an isolated singlet-like \(\mathcal X_1\), the electroweak gauge-boson rate is suppressed by \(\sin^4\theta\). The heavier, predominantly doublet-like states instead have unsuppressed weak interactions. Their contribution to singlet depletion requires rapid number-conserving conversions.

As we will see below, annihilation into dark photons alone is sufficient to deplete the symmetric component of the dark fermions.

\subsection{Condition for efficient symmetric depletion}
\label{subsec:depletion-efficiency}

Let \(Y_+\) and \(Y_-\) denote the total fermion and antifermion yields after scalar injection, with particles chosen to be the majority population. Their difference \(Y_+-Y_-=|Y_{\Delta F}^{\rm final}|\) is conserved by pair annihilation. For approximately constant entropy degrees of freedom, the asymmetric abundance equations are~\cite{Iminniyaz:2011yp}
\begin{align}
 \frac{dY_\pm}{dx}
 =-\frac{s\,\left\langle\sigma v_{\rm rel}\right\rangle_{\rm eff}}{Hx}
 \left[Y_+Y_--(Y_0^{\rm eq})^2\right],
 \label{eq:depletion-yield-equations}
\end{align}
where \(Y_0^{\rm eq}\) counts particles alone at zero chemical potential, \(\left\langle\sigma v_{\rm rel}\right\rangle_{\rm eff}\) is a single particle--antiparticle effective rate, and \(x \equiv m/T\). Once inverse pair production is negligible, integration gives the final value of \(r \equiv Y_- / Y_+\) as 
\begin{align}
 r_\infty=r_*e^{-|Y_{\Delta F}^{\rm final}| J_*}\,,
 \quad J_*\equiv\int_{x_*}^{\infty}
 \frac{s\,\left\langle\sigma v_{\rm rel}\right\rangle_{\rm eff}}{Hx}\,dx\,,
 \label{eq:depletion-residual-fraction}
\end{align}
where the matching epoch \(x_*\) corresponds to the point at which appreciable inverse production has ceased and \(r_* \equiv r(x_*)\). For an approximately constant \(s\)-wave rate during radiation domination,
\begin{align}
 J_*\simeq0.264\frac{g_{*s}}{\sqrt{g_*}}
 M_{\rm Pl}m\frac{\left\langle\sigma v_{\rm rel}\right\rangle_{\rm eff}}{x_*}.
 \label{eq:depletion-Jestimate}
\end{align}

Since \(r_*\le1\), a sufficient condition for \(r_\infty \ll 1\), \textit{e.g.}, \(r_\infty < 0.01\), is \(|Y_{\Delta F}^{\rm final}|J_*\gtrsim\ln100\). Using the asymmetry required for
\(\Omega_{\rm DM}h^2\simeq0.12\), obtained in Eq.~\eqref{eq:depletion-asymmetry-required}, yields\footnote{For a more detailed analysis, see Ref.~\cite{Graesser:2011wi}.}
\begin{align}
\left\langle\sigma v_{\rm rel}\right\rangle_{\rm eff}\gtrsim
 1\times10^{-25}~\mathrm{cm^3\,s^{-1}}
 \left(\frac{x_*}{25}\right)
 \left(\frac{g_{*s}/\sqrt{g_*}}{9}\right)^{-1}\,.
 \label{eq:depletion-cross-section}
\end{align}

If the dark-photon annihilation gives the dominant contribution to the effective annihilation rate, we have 
\begin{equation}
  \left\langle\sigma v_{\rm rel}\right\rangle_{\rm eff} \simeq \frac{1}{3} \frac{\pi\alpha_D^2}{m^2} \,,
\end{equation}
where the factor \(1/3\) is for three equally populated Dirac species. In this case, the condition~\eqref{eq:depletion-cross-section} leads to 
\begin{align}
 \alpha_D\gtrsim 0.05
 \left(\frac{m}{500~\mathrm{GeV}}\right)
 \left[\frac{x_*}{25}
 \frac{9}{g_{*s}/\sqrt{g_*}}
 \frac{0.12}{\Omega_{\rm DM}h^2}\right]^{1/2}.
 \label{eq:depletion-alpha-condition}
\end{align}
Active electroweak coannihilation reduces the required dark gauge coupling. For a light dark photon, Sommerfeld enhancement~\cite{Sommerfeld:1931qaf,Hisano:2003ec,Hisano:2004ds} and radiative bound-state formation~\cite{vonHarling:2014kha, Petraki:2015hla} can further deplete antiparticles.

\subsection{Decay of the dark photon}
\label{subsec:depletion-dark-photon-decay}

The annihilation products do not form a permanently decoupled bath
of massless dark radiation. We take
\(2m_e<m_{\gamma_D}\ll m\), so that dark photons can decay
visibly through kinetic mixing. With the hypercharge-mixing convention in Sec.~\ref{sec:model}, the low-energy electromagnetic coupling is
\(\epsilon_{\rm em}e\), where
\(\epsilon_{\rm em}\simeq\epsilon\cos\theta_W\) for
\(m_{\gamma_D}\ll m_Z\). The electron width is
\cite{Fabbrichesi:2020wbt}
\begin{align}
 \Gamma(\gamma_D\to e^+e^-)
 =\frac{\alpha\epsilon_{\rm em}^2}{3}m_{\gamma_D}
 \left(1+\frac{2m_e^2}{m_{\gamma_D}^2}\right)
 \sqrt{1-\frac{4m_e^2}{m_{\gamma_D}^2}},
 \label{eq:depletion-dark-photon-width}
\end{align}
with \(\alpha=e^2/(4\pi)\). Away from threshold and below the
muon-pair threshold, the lifetime is approximately
\begin{align}
 \tau_{\gamma_D}\simeq2.7\times10^{-3}~\mathrm{s}
 \left(\frac{\epsilon_{\rm em}}{10^{-9}}\right)^{-2}
 \left(\frac{m_{\gamma_D}}{100~\mathrm{MeV}}\right)^{-1}.
 \label{eq:depletion-dark-photon-lifetime}
\end{align}
Additional open SM channels shorten it. To avoid late electromagnetic reheating, we require the dark-photon population to disappear well before neutrino decoupling, conservatively at \(T\gtrsim5~\mathrm{MeV}\), and without appreciable entropy dilution~\cite{Ibe:2019gpv}. 

\subsection{Dark matter abundance}
\label{subsec:depletion-relic-abundance}

Adopting Eq.~\eqref{eq:ydelf}, the surviving total fermion asymmetry
is set by the CP-weighted non-thermal source,
\begin{align}
 |Y_{\Delta F}^{\rm final}|\simeq\left|\sum_i\epsilon_iY_{N_i}^{\rm NT}\right|
 =\frac32\frac{T_R}{m_\phi}
 \left|\sum_i B_i^\phi\epsilon_i\right|.
 \label{eq:depletion-source}
\end{align}
For \(r_\infty\ll1\), the observed abundance requires
\begin{align}
 |Y_{\Delta F}^{\rm final}|\simeq4.4\times10^{-13}
 \left(\frac{m}{1~\mathrm{TeV}}\right)^{-1}
 \left(\frac{\Omega_{\rm DM}h^2}{0.12}\right),
 \label{eq:depletion-asymmetry-required}
\end{align}
or equivalently
\begin{align}
\left|
\sum_i B_i^\phi\epsilon_i
\right|
&\simeq
3 \times10^{-10} 
\nonumber \\
& \times 
\left(\frac{T_R / m_\phi}{10^{-3}}\right)^{-1}
\left(\frac{\Omega_{\rm DM}h^2}{0.12}\right)
\left(\frac{m}{1~{\rm TeV}}\right)^{-1}\,.
\label{eq:required-Bepsilon}
\end{align}
If the scalar asymmetry is instead fully transferred, the source
\(|Y_{\Delta F}^{\rm final}|\) is larger by a factor of three, as we have discussed in Sec.~\ref{subsec:asym-transfer}. 

\subsection{Radiative decay of the heavier neutral fermion}
\label{subsec:depletion-radiative-decay}

The heavier neutral state can decay into the lighter state and a
photon, \(\mathcal X_2\to\mathcal X_1\gamma\), through loops involving
the charged component of the doublet and the \(W\) boson.
This decay is described by the effective transition dipole interaction
\begin{align}
\mathcal L_{\rm dip}
=\frac12\overline{\mathcal X}_1\sigma^{\mu\nu}
(\mu_{12}+i d_{12}\gamma_5)\mathcal X_2 F_{\mu\nu}
+\mathrm{h.c.},
\label{eq:depletion-transition-dipole}
\end{align}
where \(F_{\mu\nu}\) is the electromagnetic field-strength tensor and
\(\sigma^{\mu\nu}=i[\gamma^\mu,\gamma^\nu]/2\).
The corresponding decay width is
\begin{align}
\Gamma_{2\to1\gamma}
&=\frac{|\mu_{12}|^2+|d_{12}|^2}{8\pi}
\frac{(m_{\chi_2}^2-m_{\chi_1}^2)^3}{m_{\chi_2}^3}\,.
\label{eq:depletion-radiative-width}
\end{align}

To estimate the transition dipole moment, we consider the real
symmetric case in Eq.~\eqref{eq:Z-coupling-symmetric} and neglect
the small mass splittings within the loop. Rotating the doublet
dipole interaction to the mass basis yields an off-diagonal coupling
proportional to \(\sin\theta\cos\theta\). The transition dipole
moments are then approximately given by~\cite{Krall:2017xij}
\begin{align}
 |\mu_{12}|\simeq
 \frac{e g_2^2}{16\pi^2m}
 |\sin\theta\cos\theta|\,C(m)\,,
 \qquad d_{12}\simeq0\,,
 \label{eq:depletion-dipole-estimate}
\end{align}
where the loop function \(C(m)\) is approximately described by
\(C^2(m)\simeq3.2(m/\mathrm{TeV}){-0.3}\) for
\(200~\mathrm{GeV}\lesssim m\lesssim1~\mathrm{TeV}\).
Combining Eqs.~\eqref{eq:depletion-radiative-width}
and~\eqref{eq:depletion-dipole-estimate}, we obtain the lifetime 
\begin{align}
 \tau_{2\to1\gamma}\simeq 300~\mathrm{s}
 &\times C^{-2}(m)
 \left(\frac{|\sin\theta\cos\theta|}{10^{-2}}\right)^{-2}
 \nonumber\\
 &\times\left(\frac{m}{500~\mathrm{GeV}}\right)^2
 \left(\frac{\delta_\chi}{300~\mathrm{keV}}\right)^{-3} \,.
 \label{eq:depletion-chi2-lifetime}
\end{align}
Thus, \(\mathcal X_2\) cannot generally be assumed to decay before BBN.

In the chosen range \(m_{\gamma_D}>2m_e>\delta_\chi\), emission of an on-shell dark photon is forbidden. The \(e^+e^-\) and hadronic three-body channels are likewise closed. The off-shell-\(Z\) neutrino channel scales as \(G_F^2\sin^2\theta\cos^2\theta\,\delta_\chi^5\) and is subleading to the radiative channel in the doublet-loop estimate.

Despite possible lifetimes around BBN, these decays are not expected to appreciably alter primordial element abundances in the benchmark regime~\cite{Kawasaki:2020qxm}. For a nonrelativistic parent, the photon energy is
\begin{align}
 E_\gamma=\frac{m_{\chi_2}^2-m_{\chi_1}^2}{2m_{\chi_2}}
 \simeq\delta_\chi\simeq300~\mathrm{keV},
 \label{eq:depletion-photon-energy}
\end{align}
below the photodisintegration thresholds of the relevant light nuclei, including \(2.22\)~MeV for deuterium. There is no hadronic injection. Moreover, after efficient symmetric depletion, even assigning the entire DM number density to the excited state gives
\begin{align}
 \zeta_\gamma\equiv E_\gamma Y_2
 \lesssim1.3\times10^{-16}~\mathrm{GeV}
 \left(\frac{\delta_\chi}{300~\mathrm{keV}}\right)
 \left(\frac{m}{1~\mathrm{TeV}}\right)^{-1}\,,
 \label{eq:depletion-injected-energy}
\end{align}
where \(Y_2\) counts excited particles and antiparticles before their disappearance. At \(T\sim100~\mathrm{keV}\), the fractional energy release is at most of order \((4/3)(g_{*s}/g_*)\zeta_\gamma/T\sim10^{-12}\). Thus neither nonthermal nuclear destruction nor appreciable changes to the expansion rate or photon entropy are expected for these benchmarks. 

{The decay photons also have a negligible impact on the CMB spectrum for the benchmark parameters considered here. At sufficiently high redshifts, Compton scattering rapidly redistributes the injected energy among photons and electrons. However, it conserves photon number and therefore cannot generally restore a Planck spectrum on its own. Full thermalization also requires photon-number-changing processes, primarily double Compton scattering, \(e^\pm\gamma\leftrightarrow e^\pm\gamma\gamma\), and bremsstrahlung. Together, these processes efficiently erase spectral distortions at redshifts \(z\gtrsim 2\times10^6\)~\cite{Chluba:2011hw, Cyr:2026pic}, corresponding to cosmic times \(t\lesssim6\times10^6~\mathrm{s}\). The lifetime estimated in Eq.~\eqref{eq:depletion-chi2-lifetime} for our benchmark parameters is much shorter than this timescale, so the decays occur well within the era of efficient thermalization.  }

\section{Elastic scattering through dark photon exchange}
\label{sec:dark-photon-elastic}

We next consider elastic scattering of dark matter off nuclei
through dark photon exchange. Since the singlet and doublet
fermions carry the same dark charge, their coupling to the dark
photon remains diagonal and universal after rotation to the mass
basis. To leading order in kinetic mixing, the relevant
interactions are
\begin{align}
 \mathcal L \supset
 g_D\gamma_{D\mu}
 \sum_{i=1,2}\overline{\mathcal X}_i\gamma^\mu\mathcal X_i
 +\epsilon_{\rm em}e\,\gamma_{D\mu}J_{\rm em}^{\mu},
 \label{eq:elastic-dark-photon-interactions}
\end{align}
where \(J_{\rm em}^{\mu}\) is the SM electromagnetic current. With the hypercharge kinetic-mixing convention in Eq.~\eqref{eq:massYuk}, \(\epsilon_{\rm em}\simeq\epsilon\cos\theta_W\) for \(m_{\gamma_D}\ll m_Z\). Unlike the diagonal \(Z\)-boson coupling, the dark gauge coupling of \(\mathcal X_1\) is not suppressed by the singlet--doublet mixing angle. 

At momentum transfers much smaller than \(m_{\gamma_D}\),
dark photon exchange can be treated as a contact interaction.
The resulting spin-independent DM--proton cross section is
\begin{align}
 \sigma_p^{(\gamma_D)}
 &=\frac{16\pi\alpha_D\alpha\epsilon^2\cos^2\theta_W
 \,\mu_{\chi p}^2}{m_{\gamma_D}^4} 
 \label{eq:elastic-dark-photon-proton} \\ 
 &\simeq 3\times10^{-48}~\mathrm{cm^2}
 \left(\frac{\alpha_D}{0.03}\right)
 \left(\frac{\epsilon}{10^{-9}}\right)^2
 \nonumber\\
 &\times
 \left(\frac{m_{\gamma_D}}{1~\mathrm{GeV}}\right)^{-4}
 \left(\frac{\mu_{\chi p}}{0.937~\mathrm{GeV}}\right)^2\,, 
 \label{eq:elastic-dark-photon-benchmark}
\end{align}
where \(\mu_{\chi p}=m_{\chi_1}m_p/(m_{\chi_1}+m_p)\) is the DM--proton reduced mass. For the DM masses considered here, \(\mu_{\chi p}\simeq m_p\), so the cross section depends only weakly on \(m_{\chi_1}\).

At leading order, the dark photon couples to the nuclear electric
charge rather than the total nucleon number. For a nucleus with
mass \(m_A\) and atomic number \(Z\), the differential cross
section is
\begin{align}
 \frac{d\sigma_A^{(\gamma_D)}}{dE_R}
 =\frac{8\pi\alpha_D\alpha\epsilon^2\cos^2\theta_W
 \,Z^2m_A}
 {v^2\left(m_{\gamma_D}^2+2m_AE_R\right)^2}
 F_{\rm ch}^2(q),
 \label{eq:elastic-dark-photon-recoil}
\end{align}
where \(v\) is the incident DM speed, \(E_R\) is the nuclear recoil energy, \(q=\sqrt{2m_AE_R}\), and \(F_{\rm ch}(q)\) is the nuclear charge form factor, normalized by \(F_{\rm ch}(0)=1\). For \(m_{\gamma_D}=1~\mathrm{GeV}\), typical momentum transfers of tens to a hundred MeV are sufficiently small for the contact approximation to apply. In this case, the nuclear cross section is
\begin{align}
 \sigma_A^{(\gamma_D)}
 =Z^2\frac{\mu_{\chi A}^2}{\mu_{\chi p}^2}
 \sigma_p^{(\gamma_D)},
 \label{eq:elastic-dark-photon-nucleus}
\end{align}
with \(\mu_{\chi A}\) the DM--nucleus reduced mass. The coherent enhancement is therefore proportional to \(Z^2\), rather than the \(A^2\) dependence assumed in the usual isospin-conserving presentation of direct-detection limits. For a single isotope in the contact limit, the equivalent cross section in that convention is
\begin{align}
 \sigma_{N,\mathrm{equiv}}^{(\gamma_D)}
 =\left(\frac{Z}{A}\right)^2\sigma_p^{(\gamma_D)}.
 \label{eq:elastic-dark-photon-equivalent}
\end{align}
Using \(Z=54\) and \(A\simeq131\) for xenon gives
\(\sigma_{N,\mathrm{equiv}}^{(\gamma_D)}
\simeq 5\times10^{-49}~\mathrm{cm^2}\)
at the benchmark point with \(\alpha_D=0.03\).

Thus, a GeV-scale dark photon with kinetic mixing \(\epsilon=10^{-9}\) yields a small elastic-scattering cross section,\footnote{In the present model, the electroweak-doublet fermions carry both hypercharge and \(\mathrm{U}(1)_D\) charge and therefore radiatively induce a sizable kinetic mixing. The resulting low-energy value of the kinetic-mixing parameter \(\epsilon\), however, depends on the ultraviolet completion of the model, including its boundary condition at high energies and possible additional heavy states. In the following analysis, we therefore treat the low-energy value of \(\epsilon\) simply as a free parameter.} even though its coupling to dark matter is not suppressed by the singlet--doublet mixing angle. The predicted cross section at this benchmark point lies well below the current LZ limit. A lighter dark photon or larger kinetic mixing could, however, enhance the elastic-scattering cross section to a level accessible to future direct-detection experiments without substantially altering the conclusions of the preceding sections. 

\section{LZ 248~keV nuclear-recoil candidate}
\label{sec:lz}

We now examine whether the present setup can account for the LZ high-recoil event~\cite{LZ:2026axp} while remaining consistent with the stringent bounds from conventional elastic dark-matter searches.  

For \(y=\widetilde y\), the off-diagonal \(Z\)-boson coupling is proportional to \(\sin\theta\cos\theta\), whereas the diagonal coupling of the mostly singlet state is proportional to \(\sin^2\theta\).  At the nucleon level, the \(Z\)-boson coupling is dominated by the neutron vector current, since the corresponding proton coupling is suppressed by the factor \(1-4s_W^2\), with \(s_W \equiv \sin \theta_W\).  The inelastic and elastic cross sections on a neutron are therefore approximately given by\footnote{As pointed out in Refs.~\cite{Pospelov:2026ewn,Rodd:2026tyn}, the inelastic scattering cross sections quoted in Refs.~\cite{Nagata:2014wma,Nagata:2014aoa} differ from the correctly normalized result by a factor of four.  We use the corrected normalization here.}
\begin{align}
\sigma_n^{\rm inel}
&\simeq
\frac{G_F^2\mu_n^2}{2\pi}
\sin^2\theta\cos^2\theta,
\nonumber\\
\sigma_n^{\rm el}
&\simeq
\frac{G_F^2\mu_n^2}{2\pi}
\sin^4\theta,
\label{eq:inel-el-Z-cross}
\end{align}
where \(G_F\) is the Fermi constant and  \(\mu_n\equiv m_n m_{\chi_1}/(m_n+m_{\chi_1})\) is the reduced mass of the neutron \(n\) and the incoming dark matter \(\chi_1\). This hierarchy provides a simple way of obtaining an observable inelastic signal while suppressing conventional elastic scattering.\footnote{Other contributions to elastic scattering are expected to be subdominant in the parameter region of interest.  The Higgs-exchange contribution induced by the singlet--doublet mixing is strongly suppressed by the small Yukawa couplings.  The electroweak loop contribution is also negligibly small in the present case~\cite{Hisano:2011cs, Hisano:2015rsa}. }  For example, for \(m_{\chi_1}\gg m_n\), Eq.~\eqref{eq:inel-el-Z-cross} gives
\begin{align}
\sigma_n^{\rm inel}
&\simeq
7.4\times10^{-43}~{\rm cm}^2
\left(\frac{\sin\theta}{10^{-2}}\right)^2,
\nonumber\\
\sigma_n^{\rm el}
&\simeq
7.4\times10^{-47}~{\rm cm}^2
\left(\frac{\sin\theta}{10^{-2}}\right)^4 \,.
\label{eq:neutron-cross-benchmark}
\end{align}

{It is worth noting that the prediction in Eq.~\eqref{eq:inel-el-Z-cross} differs qualitatively from that for Majorana dark matter, such as the Higgsino-like neutralino. First, for Majorana dark matter, elastic scattering mediated by a vector current is absent. Inelastic scattering instead occurs between the two neutral components forming a pseudo-Dirac pair. In our model, by contrast, the inelastic scattering takes place between the neutral components of the singlet and doublet fermions, both of which are Dirac fermions. Moreover, as we discuss further at the end of this section, the predictions obtained in scenarios with mixing between singlet and doublet Majorana fermions also differ significantly from those of our model.}

To compare the elastic interaction with the conventional spin-independent limits reported by xenon experiments~\cite{LZ:2024zvo, XENON:2025vwd}, one should take into account the isospin-violating nature of the \(Z\)-boson coupling. For a nucleus with atomic number \(Z\) and mass number \(A\), the coherent elastic amplitude is proportional to
\begin{align}
  Q_W \equiv (A-Z)-\left(1-4s_W^2\right)Z \,.
  \label{eq:qw}
\end{align}
It is useful to define a xenon-equivalent isoscalar spin-independent nucleon cross section, \(\sigma_{\rm SI,eff}^{\rm Xe}\), as the cross section that would produce approximately the same xenon scattering rate under the conventional isoscalar assumption.  For a representative xenon isotope with \(A\simeq131\) and \(Z=54\), this gives
\begin{align}
\sigma_{\rm SI,eff}^{\rm Xe}
&\equiv
\sigma_n^{\rm el}
\left(\frac{Q_W}{A}\right)^2
\simeq
0.31\,\sigma_n^{\rm el}
\nonumber\\
&\simeq
2.3\times10^{-47}~{\rm cm}^2
\left(\frac{\sin\theta}{10^{-2}}\right)^4.
\label{eq:elastic-Z-estimate}
\end{align}
For dark-matter masses above a few hundred GeV, the current LZ constraint~\cite{LZ:2024zvo} on the spin-independent elastic scattering cross section can be approximately expressed as
\begin{align}
\sigma_{\rm SI,eff}^{\rm Xe}
\lesssim
3\times10^{-47}~{\rm cm}^2
\left(\frac{m_{\chi_1}}{1~{\rm TeV}}\right).
\label{eq:LZ-elastic-bound}
\end{align}
Combining this limit with Eq.~\eqref{eq:elastic-Z-estimate}, we obtain
\begin{align}
\sin\theta
\lesssim
1.1\times10^{-2}
\left(\frac{m_{\chi_1}}{1~{\rm TeV}}\right)^{1/4}.
\label{eq:mixing-LZ-bound}
\end{align}
Using this result together with Eq.~\eqref{eq:inel-el-Z-cross}, we can translate the elastic-scattering constraint into an upper bound on the \(Z\)-mediated inelastic neutron cross section,\footnote{Equivalently, defining a xenon-equivalent isoscalar nucleon cross section for the inelastic process in the same manner as \(\sigma_{\rm SI,eff}^{\rm Xe}\), the corresponding bound is
\begin{align}
\sigma_{\rm inel,eff}^{\rm Xe}
\lesssim
2.6\times10^{-43}~{\rm cm}^2
\left(\frac{m_{\chi_1}}{1~{\rm TeV}}\right)^{1/2}.
\end{align}
}
\begin{align}
\sigma_n^{\rm inel}
\lesssim
8.5\times10^{-43}~{\rm cm}^2
\left(\frac{m_{\chi_1}}{1~{\rm TeV}}\right)^{1/2}\,.
\label{eq:inelastic-bound-from-elastic}
\end{align}

The endothermic nature of the scattering naturally favors nuclear recoils at relatively high energies.  For a mass splitting \(\delta_{\chi}\equiv m_{{\chi}_2}-m_{{\chi}_1}>0\), the minimum incoming DM velocity required to produce a nuclear recoil of energy \(E_R\) is
\begin{align}
v_{\rm min}(E_R)
=
\frac{1}{\sqrt{2m_NE_R}}
\left(
\frac{m_NE_R}{\mu_{{\chi}N}}
+\delta_{\chi}
\right),
\label{eq:vmin-inelastic}
\end{align}
where \(m_N\) is the mass of the target nucleus and \(\mu_{{\chi}N}\) is the reduced mass of the incoming dark matter and the target nucleus.  As a function of the recoil energy, \(v_{\rm min}\) reaches its minimum at
\begin{align}
E_R^\star
=
\frac{\mu_{\chi N}}{m_N}\,
\delta_\chi\,,
\label{eq:LZ-recoil-peak}
\end{align}
with the corresponding minimum velocity
\begin{align}
v_{\rm min}^\star
=
\sqrt{\frac{2\delta_{\chi}}{\mu_{{\chi}N}}}.
\label{eq:vmin-star}
\end{align}
For \(m_{{\chi}_1}\simeq1~{\rm TeV}\) and a xenon target, a recoil energy \(E_R^\star\simeq250~{\rm keV}\), close to that of the LZ event, corresponds to a mass splitting of \(\delta_{\chi}\simeq 280~{\rm keV}\). The associated minimum velocity is \(v_{\rm min}^\star\simeq680~{\rm km\,s^{-1}}\), demonstrating that the signal predominantly probes the high-velocity tail of the Galactic DM distribution.

It is important to emphasize that the inelastic nucleon cross section \(\sigma_n^{\rm inel}\) in Eq.~\eqref{eq:inel-el-Z-cross} characterizes the underlying interaction strength and does not by itself determine the observable event rate.  Since the endothermic upscattering process must provide the excitation energy \(\delta_{\chi}\), only DM particles with \(v\geq v_{\rm min}(E_R)\) can contribute to a recoil of energy \(E_R\).  The differential recoil rate therefore depends on the mean inverse speed,
\begin{align}
\eta(v_{\rm min})
\equiv
\int_{|\bm v|>v_{\rm min}}
d^3\bm v\,
\frac{f_{\rm gal}(\bm v+\bm v_E)}{v},
\label{eq:halo-integral}
\end{align}
where \(\bm v\) denotes the DM velocity in the laboratory frame and \(\bm v_E\) is the velocity of the Earth with respect to the Galactic frame.  We adopt the Standard Halo Model, in which the Galactic-frame velocity distribution is taken to be a truncated Maxwell--Boltzmann distribution,
\begin{align}
f_{\rm gal}(\bm u)
=
\frac{1}{N_{\rm esc}(\pi v_0^2)^{3/2}}
e^{-u^2/v_0^2}
\Theta(v_{\rm esc}-u),
\label{eq:SHM-distribution}
\end{align}
with \(N_{\rm esc}\) chosen such that \(\int d^3\bm u\,f_{\rm gal}(\bm u)=1\), \(v_0\) the characteristic halo velocity, \(v_{\rm esc}\) the Galactic escape velocity, and \(\Theta\) the step function. 

For an endothermic mass splitting of \(\delta_{\chi}\sim{\cal O}(300~{\rm keV})\), the required \(v_{\rm min}\) lies close to the high-velocity tail of the Galactic DM distribution.  The halo integral \(\eta(v_{\rm min})\) is therefore strongly suppressed and becomes highly sensitive to \(\delta_{\chi}\) and to the assumed velocity distribution.  Consequently, the observable LZ event rate can be substantially smaller than one would infer from the nominal cross section \(\sigma_n^{\rm inel}\) alone.

\begin{figure}[t]
  \centering
  \includegraphics[width=\columnwidth]{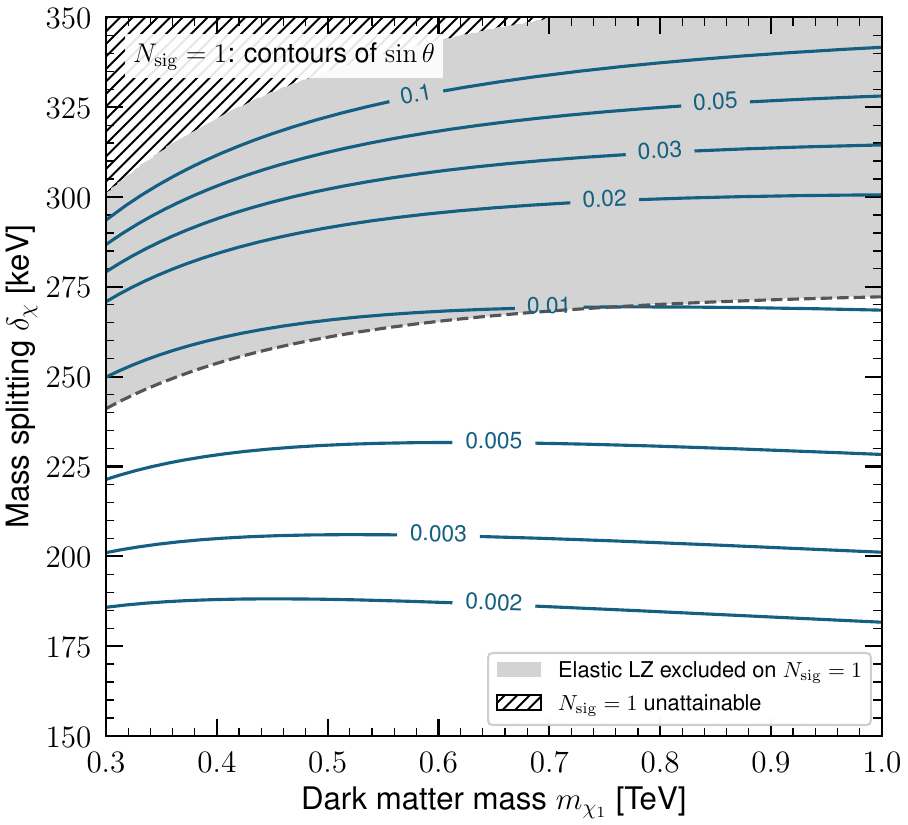}
  \caption{
    Contours of \(\sin\theta\) corresponding to one expected inelastic-scattering event in the LZ exposure of \(2.84~\mathrm{tonne\mbox{-}years}\)~\cite{LZ:2026axp}, shown in the DM mass--mass-splitting plane.  The gray-shaded region is excluded by the LZ 90\% C.L. limit on elastic scattering~\cite{LZ:2024zvo}. In the hatched region, one expected event is unattainable for any mixing angle.
  }
  \label{fig:mass_splitting}
\end{figure}

In Fig.~\ref{fig:mass_splitting}, we show contours of \(\sin\theta\) corresponding to one expected inelastic-scattering event for the LZ exposure of \(2.84~\mathrm{tonne\mbox{-}years}\)~\cite{LZ:2026axp} in the DM mass--mass-splitting plane.  The gray-shaded region is excluded by the LZ 90\% C.L. limit on elastic scattering~\cite{LZ:2024zvo}.  In the hatched region, the expected number of inelastic-scattering events remains below unity even for maximal mixing, so that one expected event cannot be obtained for any value of \(\sin\theta\). In evaluating the event rate, we assume that the entire local DM population is in the ground state \({\chi}_1\), with the excited state \({\chi}_2\) separated by a positive mass splitting \(\delta_{\chi}>0\).  Following Ref.~\cite{Baxter:2021pqo}, we adopt a local DM density of \(\rho_\chi=0.3~{\rm GeV\,cm^{-3}}\),  and use \(v_0=238~{\rm km\,s^{-1}}\), \(v_{\rm esc}=544~{\rm km\,s^{-1}}\), and \(v_E=250~{\rm km\,s^{-1}}\) for the parameters entering the velocity distribution. For the nuclear form factor, we use the Lewin--Smith parametrization~\cite{Lewin:1995rx} of the Helm form factor~\cite{Helm:1956zz},
\begin{align}
F^2(E_R)
=
\left[
\frac{3j_1(qR)}{qR}
\right]^2
e^{-q^2s^2},
\label{eq:Helm-form-factor}
\end{align}
where \(j_1\) is the spherical Bessel function of the first kind, \(q=\sqrt{2m_NE_R}\), \(R^2 =c^2+{7\pi^2}a^2/3-5s^2\), \(s=0.9~{\rm fm}\), \(a =0.52~{\rm fm}\), \(c =(1.23A^{1/3}-0.60)~{\rm fm}\). The recoil-energy-dependent detection efficiency is extracted from Fig.~S2 of Ref.~\cite{LZ:2026axp} and is included in computing the expected number of events.

As shown in Fig.~\ref{fig:mass_splitting}, there exists a sizable region of parameter space in which the LZ high-recoil event can be accommodated while satisfying the stringent constraint from elastic scattering.  Over the range of DM masses displayed in the figure, this is possible for mixing angles of \(\sin\theta\lesssim 10^{-2}\) and mass splittings \(\delta_{\chi}\lesssim 250~{\rm keV}\). We, however, stress that the precise contours shown in Fig.~\ref{fig:mass_splitting} are subject to astrophysical and experimental uncertainties.  In particular, because endothermic scattering with a mass splitting of a few hundred keV probes the high-velocity tail of the Galactic DM distribution, the predicted event rate can depend sensitively on the assumed halo model, including the velocity distribution and the values of \(v_0\), \(v_{\rm esc}\), and \(v_E\).  Possible departures from the Standard Halo Model can therefore shift the preferred values of \(\delta_{\chi}\) and \(\sin\theta\). {We note that Fig.~\ref{fig:mass_splitting} merely shows the parameter region that reproduces the observed number of events, without taking into account the resulting recoil-energy spectrum. Although a mass splitting of \(\simeq  250~\mathrm{keV}\) roughly reproduces the recoil energy of the LZ event through Eq.~\eqref{eq:LZ-recoil-peak}, such a mass splitting may be in tension with the absence of events at lower recoil energies. On the other hand, as discussed below, the absence of events in the high-energy sideband of the LZ search may favor a somewhat smaller mass splitting. A more precise determination of the viable parameter region would require a dedicated likelihood analysis incorporating astrophysical uncertainties as well as the detailed LZ detector response.}

An important feature of our scenario is that the DM abundance is determined by the primordial dark-sector asymmetry rather than by thermal freeze-out.  The DM mass is therefore not fixed by the requirement of reproducing the observed relic abundance through annihilation.  In particular, the DM can be substantially lighter than \(1~{\rm TeV}\) while still accounting for the entire observed DM density, with the required abundance obtained by an appropriate value of the dark-sector asymmetry.  This is in sharp contrast to a thermal WIMP with fixed electroweak interactions, such as a Higgsino, for which lowering the DM mass below its thermal value leads to an underabundant relic.

{
The freedom in the DM mass, together with the smaller mass splittings favored in our model, is potentially advantageous in addressing the high-energy sideband of the LZ search. Ref.~\cite{Rodd:2026tyn} pointed out that an approximately \(1~\mathrm{TeV}\) inelastic DM candidate capable of explaining the observed event tends to predict additional recoils at higher energies, whereas no events were observed in the corresponding high-energy \(\mathrm{S}1c\) sideband. This creates a potential tension for the thermal Higgsino interpretation, although its quantitative significance depends on the detector acceptance in the sideband region, which is not publicly available, and on the assumed Galactic DM velocity distribution~\cite{Rodd:2026tyn}. Lighter DM can alleviate this tension because its smaller available kinetic energy limits the high-energy recoil tail. Indeed, Ref.~\cite{Rodd:2026tyn} discussed that a nonthermal Higgsino with a mass around \(500~\mathrm{GeV}\) can evade the sideband constraint. In our asymmetric scenario, such lighter DM can account for the full observed abundance because its relic density is determined by the primordial asymmetry. Furthermore, the smaller mass splittings favored in our model lower the characteristic recoil-energy scale \(E_R^\star=\mu_{\chi N}\delta_\chi/m_N\) (see Eq.~\eqref{eq:LZ-recoil-peak}). After normalization to the observed event, this may further reduce the relative population of high-energy recoils. These features make our scenario promising in light of the sideband constraint, although a quantitative assessment requires the full recoil spectrum folded with the detector response.
}

Our scenario also leads to several characteristic predictions.  First and foremost, the elastic and inelastic scattering cross sections are correlated because they are controlled by the same mixing angle.  In the small-mixing limit, Eq.~\eqref{eq:neutron-cross-benchmark} implies
\begin{align}
\sigma_n^{\rm el}
\simeq
7.4\times10^{-47}~{\rm cm}^2
\left(
\frac{\sigma_n^{\rm inel}}
{7.4\times10^{-43}~{\rm cm}^2}
\right)^2.
\label{eq:pred}
\end{align}
Thus, an inelastic scattering rate sufficiently large to account for the LZ event necessarily implies a nonzero elastic scattering cross section.  In particular, for values of \(\sin\theta\) close to but below the current elastic-scattering bound shown in Fig.~\ref{fig:mass_splitting}, the predicted elastic cross section lies within reach of future direct-detection experiments.  The correlated observation of elastic and inelastic scattering satisfying Eq.~\eqref{eq:pred} would therefore provide a distinctive test of the present scenario. In addition, the inelastic nature of the scattering leads to distinctive astrophysical and target-dependent signatures.  Since endothermic scattering with \(\delta_{\chi}\sim{\cal O}(100)~{\rm keV}\) probes the high-velocity tail of the Galactic DM distribution, a relatively large annual modulation of the event rate is expected~\cite{Tucker-Smith:2004mxa}.  The scattering rate also exhibits a pronounced dependence on the target nucleus.  First, the kinematic threshold depends on the DM--nucleus reduced mass, favoring heavy nuclei such as xenon and tungsten, while scattering on lighter nuclei can become kinematically inaccessible for the mass splittings relevant to the LZ event~\cite{Tucker-Smith:2004mxa, Su:2026rwz}.  Second, in our scenario the scattering is mediated by \(Z\)-boson exchange and is therefore not isospin symmetric.  In particular, the effective proton coupling is suppressed by a factor \(|1-4s_W^2|\simeq 0.075\) compared to the effective neutron coupling, and the coherent nuclear amplitude is proportional to the weak charge in Eq.~\eqref{eq:qw}, rather than simply to the mass number \(A\).  The interaction is thus predominantly sensitive to the neutron content of the target nucleus.  This provides an additional, characteristic target dependence beyond that arising from the endothermic kinematics~\cite{Feng:2011vu}.  Consequently, if the LZ event originates from the mechanism proposed here, experiments employing different nuclear targets should observe correlated rates determined by both their inelastic kinematics and their weak charges.  Comparing signals in xenon with those in other heavy-target experiments would therefore provide an important test of the \(Z\)-mediated inelastic-scattering interpretation. 

Finally, the present setup predicts a nearly degenerate electroweak doublet of fermions associated with the DM state. After the electroweak-symmetry breaking, electroweak radiative corrections lift the degeneracy between the charged and neutral doublet-like states, making the charged state heavier by \(\sim 350~{\rm MeV}\)~\cite{Thomas:1998wy}.  The charged state then dominantly decays into the neutral doublet-like state and a charged pion, with a decay length of \(\sim 1~{\rm cm}\).  Such a relatively long-lived charged particle gives rise to a short disappearing-track and can be searched for at future LHC runs with dedicated disappearing-track analyses~\cite{Fukuda:2017jmk, ATLAS:2022rme, CMS:2023mny}. Moreover, future high-precision electroweak measurements at \(e^+e^-\) circular colliders, such as FCC-ee, can probe such a vector-like electroweak doublet for masses \(\lesssim 500~{\rm GeV}\)~\cite{Maura:2024zxz, Nagata:2025ycf, Greljo:2025ggc, Baer:2025tge, Hamaguchi:2026iry}. 

Before closing this section, we comment on the relation of our scenario to the recent singlet--doublet interpretations proposed in Refs.~\cite{Borah:2026zwf,Lee:2026jxl}.  As in our setup, these models introduce mixing between an electroweak-singlet fermion and a vector-like \(\mathrm{SU}(2)_L\) doublet, thereby suppressing the \(Z\)-mediated interaction of the mostly singlet DM state.  There are, however, two important differences.  First, Refs.~\cite{Borah:2026zwf,Lee:2026jxl} introduce Majorana mass terms that split the light Dirac state into a pair of Majorana fermions.  The Majorana nature of the DM eliminates the diagonal vector coupling to the \(Z\) boson and leaves an off-diagonal coupling between the two Majorana states.  At the same time, however, the DM remains capable of pair annihilation in the present Universe.  Consequently, solar capture can lead to DM annihilation and an associated high-energy neutrino signal, so that constraints from IceCube remain relevant; this constraint was explicitly studied in Ref.~\cite{Lee:2026jxl}.  In our setup, by contrast, no Majorana mass term is introduced and the dark fermion remains Dirac.  The conserved dark-sector charge therefore preserves the primordial asymmetry, and the antiparticle population is strongly depleted at late times.  As a result, captured DM particles in the Sun have essentially no antiparticles with which to annihilate, allowing our scenario to evade the corresponding IceCube constraint. Second, the dependence of the inelastic scattering rate on the singlet--doublet mixing angle is parametrically different.  In Refs.~\cite{Borah:2026zwf,Lee:2026jxl}, the off-diagonal \(Z\) coupling between the two Majorana states is proportional to \(\sin^2\theta\), and hence the inelastic scattering cross section scales as \(\sigma^{\rm inel}\propto\sin^4\theta\).  In our case, the transition occurs between two distinct Dirac mass eigenstates, for which the off-diagonal \(Z\) coupling is proportional to \(\sin\theta\cos\theta\).  The corresponding cross section therefore scales as \(\sigma^{\rm inel}\propto\sin^2\theta\cos^2\theta\simeq\sin^2\theta\) in the small-mixing limit.  This weaker mixing suppression allows the LZ event to be accommodated with comparatively small mixing angles, \(\sin\theta\lesssim{\cal O}(10^{-2})\), while simultaneously satisfying the stringent constraint on elastic scattering.  Thus, despite the apparent similarity of the underlying singlet--doublet mixing structures, the asymmetric Dirac nature of DM in our model leads to qualitatively different direct- and indirect-detection phenomenology from the Majorana scenarios of Refs.~\cite{Borah:2026zwf,Lee:2026jxl}.

\section{Discussion}
\label{sec:discussion}
In this work, we have proposed an asymmetric inelastic dark-matter scenario motivated by the high-energy nuclear-recoil candidate recently reported by LZ. The central idea is that a particle--antiparticle asymmetry and inelastic electroweak scattering provide complementary ingredients for explaining the event. The dark-matter asymmetry suppresses annihilation after capture in the Sun and thereby avoids the stringent IceCube constraint that challenges the thermal Higgsino interpretation~\cite{Pospelov:2026ewn,DiMauro:2026dqp,Bose:2026ndd,Nguyen:2026lui}, while singlet--doublet mixing generates the \(Z\)-mediated inelastic interaction required for the high-energy recoil.

We have presented a simple realization in which the dark matter is a predominantly electroweak-singlet Dirac fermion that mixes weakly with a nearly degenerate vector-like electroweak doublet. The dark-sector asymmetry is generated by CP-violating decays of heavy Majorana fermions produced nonthermally through inflaton decays, in close analogy with nonthermal leptogenesis. The symmetric component can subsequently be depleted through dark-sector interactions, leaving an asymmetric relic. Since the relic abundance is determined by the primordial asymmetry rather than thermal freeze-out, the dark-matter mass is not fixed to the thermal value characteristic of an electroweak multiplet. In particular, masses substantially below \(1~\mathrm{TeV}\) can account for the full dark-matter abundance, which can also alleviate the possible tension with the absence of events in the higher-energy LZ sideband~\cite{Rodd:2026tyn}.

A key feature of the model is the different mixing-angle dependence of the diagonal and off-diagonal \(Z\) couplings. In the small-mixing limit, the elastic and inelastic cross sections scale as \(\sin^4\theta\) and \(\sin^2\theta\), respectively. This allows an observable inelastic signal to coexist with stringent elastic-scattering limits. We find a sizable region of parameter space with \(\sin\theta\lesssim10^{-2}\) and a mass splitting of a few hundred keV in which the LZ event can be accommodated. The scenario is testable through the correlated elastic and inelastic scattering rates, annual modulation and target dependence characteristic of endothermic \(Z\)-mediated scattering, and complementary searches for the electroweak-doublet states at colliders and future precision experiments. Dark-photon-mediated elastic scattering provides an additional probe whose importance depends on the dark-photon mass and kinetic mixing.

The precise parameter region compatible with the LZ event remains subject to astrophysical and experimental uncertainties, since scattering with a mass splitting of a few hundred keV is highly sensitive to the high-velocity tail of the Galactic dark-matter distribution. A dedicated likelihood analysis including halo uncertainties and the detailed detector response would therefore be important if the high-energy excess persists with additional exposure. Nevertheless, our results demonstrate that the IceCube constraint need not disfavor an inelastic interpretation of the LZ event itself. If the event is ultimately associated with dark matter, the absence of a corresponding solar-neutrino signal may instead point toward an asymmetric, rather than thermal, origin of the dark-matter relic abundance.

\begin{acknowledgments}
N.N. thanks Satoshi Shirai for valuable discussions on various aspects of the LZ event shortly after its announcement.
T.T.Y. also thanks Masahiro Ibe, Keith Olive and Satoshi Shirai for discussion on the IceCube neutrino flux problem, and  Jie Sheng for discussion on a possible lower bound of $\delta_\chi$. The work of N.N. was supported in part by the Grant-in-Aid for Scientific Research C (No. 25K07314). T.~T.~Y. is supported by MEXT Grant No.~24H02244 and the World Premier International Research Center Initiative (WPI), MEXT, Japan (Kavli IPMU).

\end{acknowledgments}

\bibliographystyle{name}
\bibliography{references}

\end{document}